\documentclass[11pt,showkeys,aps,pra]{revtex4}
\usepackage[utf8x]{inputenc}
\usepackage{ucs}

\usepackage{amsmath,amssymb}
\usepackage{latexsym}
\usepackage{amsfonts}
\usepackage{dcolumn}% Align table columns on decimal point
\usepackage{bm}% bold math
\usepackage[usenames]{color}
\usepackage{multirow}
\usepackage{caption}
\usepackage{graphicx}
\usepackage{hyperref}
\usepackage{subfig}
\usepackage{booktabs}
\usepackage{xcolor}
\usepackage[normalem]{ulem}
\usepackage{float} % Allows putting an [H] in \begin{figure} to specify the exact location of the figure
\usepackage{wrapfig} % Allows in-line images such as the example fish picture
\usepackage{upgreek} % para poner letras griegas sin cursiva.
\usepackage{cancel} % para tachar.
\usepackage{mathdots} % para el comando \iddots
\usepackage{mathrsfs} % para formato de letra
\graphicspath{{figures/}} % Specifies the directory where pictures are stored

\begin{document}

\title{Exact Shannon entropies for the multidimensional harmonic states}

%\affiliation{}

\author{I.V. Toranzo}
\email[]{ivtoranzo@ugr.es}
\affiliation{Departamento de Matem\'atica Aplicada, Universidad Rey Juan Carlos, 28933 Madrid, Spain}
\affiliation{Instituto Carlos I de F\'{\i}sica Te\'orica y Computacional, Universidad de Granada, Granada 18071, Spain}

%\author{D. Puertas-Centeno}
%\email[]{vidda@correo.ugr.es}
%\affiliation{Departamento de F\'{\i}sica At\'{o}mica, Molecular y Nuclear, Universidad de Granada, Granada 18071, Spain}
%\affiliation{Instituto Carlos I de F\'{\i}sica Te\'orica y Computacional, Universidad de Granada, Granada 18071, Spain}

\author{J.S. Dehesa}
\email[]{dehesa@ugr.es}
\affiliation{Departamento de F\'{\i}sica At\'{o}mica, Molecular y Nuclear, Universidad de Granada, Granada 18071, Spain}
\affiliation{Instituto Carlos I de F\'{\i}sica Te\'orica y Computacional, Universidad de Granada, Granada 18071, Spain}

\begin{abstract}
In this work we determine and discuss the entropic uncertainty measures of Shannon type for all the discrete stationary states of the multidimensional harmonic systems directly in terms of the states' hyperquantum numbers, the dimensionality and the oscillator strength.  We have found that these entropies have a monotonically increasing behavior when both the dimensionality and the population of the states are increasing
\end{abstract}

%\pacs{}

\keywords{Shannon entropy; multidimensional harmonic systems}
\maketitle

\section{Introduction}

The Shannon information entropy of a probability distribution \cite{shannon,shannon2} is well known to be not only the cornerstone of the classical communication and computation \cite{cover,luenberger} but also to play a relevant role for the analysis, description and interpretation of numerous physical phenomena and systems in a great variety of scientific fields ranging from electronic structure \cite{sen,KDsen,nagy,site}), atomic spectroscopy \cite{rosario1,He,toranzo2016} and molecular systems \cite{nalewajski,dehesa2006,esquivel,esquivel2015,esquivel2016} till quantum uncertainty \cite{rudnicki,rudnicki2}, relativistic physics \cite{wan}, seismic events \cite{telesca}, signals and chaos theory \cite{vignat,rosso,dehesa2002}, neural networks \cite{yarrow}, applied mathematics \cite{dehesa2001,dehesa2007}, quantum information \cite{nielsen,holik} and Bose-Einstein condensates \cite{sriraman}.\\

The Shannon entropy of the probability density $\rho(\mathbf{x}), \mathbf{x} = (x_1, ..., x_D)$, which characterizes the physical states of a $D$-dimensional quantum system is defined as
\begin{equation}
\label{se}
S[\rho] = - \int_{\mathbb{R}_D}\rho(\mathbf{x})\ln\rho(\mathbf{x})d\mathbf{x} \, .
\end{equation}
This quantity, which measures the total spreading of the density, gives a much more appropriate definition of the position uncertainty than the standard-deviation-based Heisenberg measure. This is basically because the latter depends on a specific point of the domain of the density and it gives a large weight to the tails of the distribution (see e.g., \cite{hilge}), what is only true for some particular distributions such as those which fall off exponentially.\\

The exact determination of the Shannon entropy of multidimensional quantum systems is an impossible task because the associated Schr\"odinger equation is not analytically solvable except for a small bunch of elementary systems subject to a quantum-mechanical potential of zero-range, Coulomb or harmonic (i.e., oscillator-like) type. Presently there exists a huge interest for $D$-dimensional harmonic oscillators in general quantum mechanics \cite{gallup,chang,dong2011,adegoke16,adegoke16bis,adegoke17,buyukasik,armstrong3,armstrong4,dean}, quantum chromodynamics and elementary particle physics \cite{hoof,bures}, heat transport \cite{asadian,lepri}, information theory \cite{yanez1994,roven,aptekarev2016,assche95}, fractional thermostatistics \cite{roven} and quantum information \cite{plenio,galve,zeilinger}. Even for these systems the analytical evaluation of the Shannon entropy for their stationary states is a formidable task because it is a logarithmic integral functional of some special functions of applied mathematics and mathematical physics \cite{nikiforov,dehesa2001} (e.g., orthogonal polynomials, hyperspherical harmonics,...) which control the associated wavefunctions. This issue is much more difficult than the determination of other uncertainty measures like the Heisenberg-like and Fisher information ones, not only analytically but also numerically. The latter is basically because a naive numerical evaluation using quadratures is not convenient due to the increasing number of integrable singularities when the principal hyperquantum number is increasing, which spoils any attempt to achieve reasonable accuracy even for rather small hyperquantum number \cite{buyarov}. \\

In this work we determine in a closed form the Shannon entropy for all the discrete stationary states of the $D$-dimensional harmonic oscillator system (i.e., a particle moving under the action of a quadratic potential) from first principles, i.e. directly in terms of the hyperquantum numbers of the states and the oscillator strength. Up until now this quantity has been computed only for the one-dimensional harmonic states \cite{jordi} and for the high-energy (i.e., Rydberg) states of the two- and $D$-dimensional harmonic systems 
 \cite{dehesa1998,aptekarev2016,dehesa2017}. In addition, the Shannon entropy for the high-dimensional (i.e., pseudoclassical) harmonic states has been already conjectured \cite{puertas2018b}. We should also mention here that the other basic uncertainty measures (namely the Heisenberg-like measures, the Fisher information and the R\'enyi entropies) have been recently calculated for all stationary states of the multidimensional harmonic system in \cite{ray,zozor,suslov}, \cite{romera2005} and \cite{puertas2018b} respectively, for the high-dimensional harmonic states in \cite{puertas2018} and for the Rydberg harmonic states in \cite{aptekarev2016,dehesa2017}. See also \cite{ADST} for the values of the R\'enyi entropies for the Rydberg states of the one-dimensional harmonic oscillator. \\

In the following we first give the probability densities which characterize the stationary states of the multidimensional harmonic  oscillator system. Then, we calculate the Shannon entropy of these densities in an explicit manner.\\

\section{Quantum probability densities of harmonic states}

The Schr\"{o}dinger equation of the $D$-dimensional harmonic oscillator has the form \cite{yanez1994} (see also \cite{gallup,kostelecky,lanfear,majernik})
\begin{equation}\label{schrodinger}
\left( -\frac{1}{2} \vec{\nabla}^{2}_{D} + V(r)\right) \Psi \left( \vec{r} \right) = E \Psi \left(\vec{r} \right),
\end{equation}
in atomic units, where $\vec{\nabla}_{D}$ and $V(r)$ denote the $D$-dimensional gradient operator and the quadratic potential $V(\vec{r}) = \frac{1}{2}k(x_{1}^{2}+\ldots + x_{D}^{2}) \equiv \frac{1}{2}kr^{2}$, respectively. It is known that in Cartesian coordinates the stationary states ${\{n_i; i = 1,\ldots, D\}}$ of this system are characterized by the energy eigenvalues \cite{yanez1994} (see also \cite{gallup,kostelecky,lanfear,majernik}) 
%\begin{equation}
%V(\vec{r}) = \frac{1}{2}k(x_{1}^{2}+\ldots + x_{D}^{2}) = \frac{1}{2}kr^{2}.
%\end{equation}
\begin{equation}
\label{HOEL}
E_{\{n_i\}} = \left(N + \frac{D}{2}\right)  \omega,
\end{equation}
with 
\[
\omega = \sqrt{k}, \quad N = \sum_{i=1}^{D}n_{i} \quad \text{with} \quad n_{i}=0,1,2,\ldots
\]
and the associated eigenfunctions \cite{adegoke16,adegoke17}
\begin{equation}
\label{HOEF}
\psi_{\{n_i\}}(\vec{r}) = \mathcal{N}\, e^{-\frac{1}{2}\alpha(x_{1}^{2}+x_{2}^{2}+\ldots+x_{D}^{2})}H_{n_{1}}(\sqrt{\alpha}\, x_{1})\cdots H_{n_{D}}(\sqrt{\alpha}\, x_{D}), \quad \alpha = \sqrt{k}
\end{equation}	
where the normalization constant
\[
\mathcal{N} = \frac{1}{\sqrt{2^{N}n_{1}!n_{2}!\cdots n_{D}! }}\left(\frac{\alpha}{\pi}\right)^{D/4}
\]
and $\{H_{n_i}(x);i = 1,\ldots, D\}$ denote the Hermite polynomials of degree $n_i$ orthogonal with respect the weight function $\omega_{n}(x) = e^{-x^{2}}$ in $(-\infty, \infty)$.\\
Then, the corresponding quantum probability densities are given by
\begin{equation}
\label{HOPD}
\rho_{\{n_i\}}(\vec{r}) = |\psi_{\{n_i\}}(\vec{r})|^{2} = \mathcal{N}^{2} e^{-\alpha(x_{1}^{2}+x_{2}^{2}+\ldots+x_{D}^{2})}H_{n_{1}}^{2}(\sqrt{\alpha}\, x_{1})\cdots H_{n_{D}}^{2}(\sqrt{\alpha}\, x_{D})
\end{equation}
and
\begin{equation} 
\gamma_{\{n_i\}}(\vec{p}) = |\tilde{\psi}_{\{n_i\}}(\vec{p})|^{2} = \frac{1}{\alpha^{D}}\rho_{\{n_i\}}\left(\frac{\vec{p}}{\alpha}\right)
\end{equation}
in the position and momentum spaces, respectively. The symbol $\tilde{\psi}_{\{n_i\}}(\vec{p})$ denotes the momentum eigenfunction of the system, i.e., the Fourier transform of the position eigenfunction $\psi_{\{n_i\}}(\vec{r})$.
%\begin{align}
%\label{HOMPD}
%\gamma_{N}(\vec{p}) &= |\tilde{\psi}_{N}(\vec{p})|^{2} = \mathcal{\tilde{N}}^{2} e^{-\frac{1}{\alpha}(p_{1}^{2}+p_{2}^{2}+\ldots+p_{D}^{2})}H_{n_{1}}^{2}\left(\frac{ p_{1}}{\sqrt{\alpha}}\right)\cdots H_{n_{D}}^{2}\left(\frac{ p_{D}}{\sqrt{\alpha}}\right),\nonumber \\
%&= \rho_{N}\left(\frac{\vec{p}}{\alpha}\right)
%\end{align}
%where the normalization constant is 
%\[
%\mathcal{\tilde{N}}  = \frac{1}{\sqrt{2^{N}n_{1}!n_{2}!\cdots n_{D}! }}\left(\frac{1}{\pi\alpha}\right)^{D/4}.
%\]
%Thus, the Rényi entropy of the harmonic oscillator in cartesian coordinates can be written as
%\begin{align}
%\label{HORE}
%R_{q}[\rho_{N}] &= \frac{1}{1-q}\log \int_{-\infty}^{\infty} dx_{1}\ldots \int_{-\infty}^{\infty} dx_{D} \, [\rho_{N}(\vec{r})]^{q} \nonumber \\
%& =  \frac{1}{1-q}\log\left( \mathcal{N}^{2q}\int_{-\infty}^{\infty} e^{-\alpha q x_{1}^{2}}H_{n_{1}}^{2q}(\sqrt{\alpha}\, x_{1}) \, dx_{1} \cdots \int_{-\infty}^{\infty} e^{-\alpha q x_{D}^{2}}H_{n_{D}}^{2q}(\sqrt{\alpha}\, x_{D})\, dx_{D} \right)
%\end{align}

\section{Shannon entropy of harmonic states}

Here we calculate the Shannon entropy (\ref{HOEL}) of the multidimensional harmonic probability density (\ref{HOPD}), that is,
\begin{align}
\label{cchose1}
S[\rho_{\{n_i\}}] &= - \int_{\mathbb{R}^{D}} \rho_{\{n_i\}}(x_{1},\ldots,x_{D})\log [\rho_{\{n_i\}}(x_{1},\ldots,x_{D})] \, dx_{1}\cdots dx_{D} \nonumber \\
&= -\mathcal{N}^{2}\int_{\mathbb{R}^{D}}e^{-\alpha(x_{1}^{2}+x_{2}^{2}+\ldots+x_{D}^{2})}H_{n_{1}}^{2}(\sqrt{\alpha}\, x_{1})\cdots H_{n_{D}}^{2}(\sqrt{\alpha}\, x_{D})\times \nonumber \\ 
&  \times \log\left[\mathcal{N}^{2} e^{-\alpha(x_{1}^{2}+x_{2}^{2}+\ldots+x_{D}^{2})}H_{n_{1}}^{2}(\sqrt{\alpha}\, x_{1})\cdots H_{n_{D}}^{2}(\sqrt{\alpha}\, x_{D}) \right]\, dx_{1}\cdots dx_{D} \nonumber \\
&= \mathcal{I}_{1}^{(D)} +   \mathcal{I}_{2}^{(D)}  +  \mathcal{I}_{3}^{(D)},
\end{align}
where the integrals $\mathcal{I}_{1}^{(D)}$, $\mathcal{I}_{2}^{(D)}$ and $\mathcal{I}_{3}^{(D)}$ are given by
\begin{equation}
\mathcal{I}_{1}^{(D)} = -\mathcal{N}^{2}\log\left(\mathcal{N}^{2}\right)\int_{\mathbb{R}^{D}} e^{-\alpha(x_{1}^{2}+x_{2}^{2}+\ldots+x_{D}^{2})}\,\Bigg[\Pi_{i=1}^{D}H_{n_{i}}^{2}(\sqrt{\alpha}\, x_{i})\Bigg]\, dx_{1}\cdots dx_{D},
\end{equation}
\begin{equation}
\mathcal{I}_{2}^{(D)}=\alpha\,\mathcal{N}^{2}\int_{\mathbb{R}^{D}} e^{-\alpha(x_{1}^{2}+x_{2}^{2}+\ldots+x_{D}^{2})}\,\Bigg[\Pi_{i=1}^{D}H_{n_{i}}^{2}(\sqrt{\alpha}\, x_{i})\Bigg]\,(x_{1}^{2}+x_{2}^{2}+\ldots+x_{D}^{2})\, dx_{1}\ldots dx_{D} \end{equation}
and
\begin{equation}
\mathcal{I}_{3}^{(D)}= -\mathcal{N}^{2}\int_{\mathbb{R}^{D}} e^{-\alpha(x_{1}^{2}+x_{2}^{2}+\ldots+x_{D}^{2})}\,\Bigg[\Pi_{i=1}^{D}H_{n_{i}}^{2}(\sqrt{\alpha}\, x_{i})\Bigg]\,\log\,\Bigg[\Pi_{i=1}^{D}H_{n_{i}}^{2}(\sqrt{\alpha}\, x_{i})\Bigg]\, dx_{1}\cdots dx_{D},
\end{equation}
respectively, whose computation is shown in the following. The first integral $\mathcal{I}_{1}^{(D)}$ can be calculated as
\begin{align}
\label{I11}
\mathcal{I}_{1}^{(D)} &= -\mathcal{N}^{2}\log\left(\mathcal{N}^{2}\right)\int_{\mathbb{R}^{D}} e^{-\alpha(x_{1}^{2}+x_{2}^{2}+\ldots+x_{D}^{2})}\,\Bigg[\Pi_{i=1}^{D}H_{n_{i}}^{2}(\sqrt{\alpha}\, x_{i})\Bigg]\, dx_{1}\cdots dx_{D} \nonumber \\
&= -\mathcal{N}^{2}\log\left(\mathcal{N}^{2}\right) \Pi_{i=1}^{D} \int_{-\infty}^{\infty} e^{-\alpha x_{i}^{2}}H_{n_{i}}^{2}(\sqrt{\alpha}x_{i}) \, dx_{i} \nonumber \\
&= -\mathcal{N}^{2}\log\left(\mathcal{N}^{2}\right) \Pi_{i=1}^{D} \left(\alpha^{-\frac{1}{2}}2^{n_{i}}n_{i}!\sqrt{\pi} \right)\nonumber \\
%&= -\mathcal{N}^{2}\log\left(\mathcal{N}^{2}\right) \left(\frac{\pi}{\alpha}\right)^{\frac{D}{2}}2^{n_{1} +\ldots + n_{D}}n_{1}!\cdots n_{D}!\nonumber \\
%&= -\log\mathcal{N}^{2}\nonumber \\
&=  \left(\sum_{i=1}^{D}n_{i}\right)\log 2 + \sum_{i=1}^{D} \log (n_{i}!) + \frac{D}{2} \log \left(\frac{\pi}{\alpha}\right).
\end{align}
Then, the second integral $\mathcal{I}_{2}^{(D)}$ can be determined as
\begin{align}
\label{I21}
\mathcal{I}_{2}^{(D)} &=\alpha\,\mathcal{N}^{2}\int_{\mathbb{R}^{D}} e^{-\alpha(x_{1}^{2}+x_{2}^{2}+\ldots+x_{D}^{2})}\,\Bigg[\Pi_{i=1}^{D}H_{n_{i}}^{2}(\sqrt{\alpha}\, x_{i})\Bigg]\,(x_{1}^{2}+x_{2}^{2}+\ldots+x_{D}^{2})\, dx_{1}\ldots dx_{D} \nonumber \\
&= \alpha \mathcal{N}^{2}\,\Pi_{i=1}^{D}\int_{\mathbb{R}^{D}} x_{i}^{2}\,e^{-\alpha(x_{1}^{2}+x_{2}^{2}+\ldots+x_{D}^{2})}H_{n_{1}}^{2}(\sqrt{\alpha}\, x_{1})\cdots H_{n_{D}}^{2}(\sqrt{\alpha}\, x_{D})\, dx_{1}\ldots dx_{D}\nonumber \\ 
%&= \alpha \mathcal{N}^{2}\Bigg[\int_{\mathbb{R}^{D}} x_{1}^{2}\,e^{-\alpha(x_{1}^{2}+x_{2}^{2}+\ldots+x_{D}^{2})}H_{n_{1}}^{2}(\sqrt{\alpha}\, x_{1})\cdots H_{n_{D}}^{2}(\sqrt{\alpha}\, x_{D})\, dx_{1}\ldots dx_{D}   + \cdots \nonumber \\
& = \alpha \mathcal{N}^{2} \int_{-\infty}^{\infty} x_{1}^{2}\,e^{-\alpha x_{1}^{2}}H_{n_{1}}^{2}(\sqrt{\alpha}\, x_{1}) \, dx_{1} \,\Bigg[\Pi_{i=2}^{D}\int_{-\infty}^{\infty} e^{-\alpha x_{i}^{2}}H_{n_{i}}^{2}(\sqrt{\alpha}\, x_{i}) \, dx_{i}\Bigg] + \ldots\nonumber \\
& + \int_{-\infty}^{\infty} x_{D}^{2}\, e^{-\alpha x_{D}^{2}}H_{n_{D}}^{2}(\sqrt{\alpha}\, x_{D}) \, dx_{D}\,\Bigg[\Pi_{i=1}^{D-1}\int_{-\infty}^{\infty} e^{-\alpha x_{i}^{2}}H_{n_{i}}^{2}(\sqrt{\alpha}\, x_{i}) \, dx_{i}\Bigg]\nonumber \\
& = \alpha \mathcal{N}^{2}\Bigg[ \left(\frac{\pi}{\alpha}\right)^{\frac{D-1}{2}}2^{n_{2}+\ldots+n_{D}}n_{2}!\cdots n_{D}! \int_{-\infty}^{\infty} x_{1}^{2}\,e^{-\alpha x_{1}^{2}}H_{n_{1}}^{2}(\sqrt{\alpha}\, x_{1}) \, dx_{1} + \ldots\nonumber \\
& + \left(\frac{\pi}{\alpha}\right)^{\frac{D-1}{2}}2^{n_{1}+\ldots+n_{D-1}}n_{1}!\cdots n_{D-1}! \int_{-\infty}^{\infty} x_{D}^{2}\, e^{-\alpha x_{D}^{2}}H_{n_{D}}^{2}(\sqrt{\alpha}\, x_{D}) \, dx_{D} \Bigg]\nonumber \\
& = \alpha \mathcal{N}^{2}\left(\frac{\pi}{\alpha}\right)^{\frac{D-1}{2}}2^{N}\Bigg(\Pi_{i=1}^{D}n_{i}!\Bigg) \Bigg[\Pi_{i=1}^{D} \frac{1}{2^{n_{i}}n_{i}!}\int_{-\infty}^{\infty} x_{i}^{2}\,e^{-\alpha x_{i}^{2}}H_{n_{i}}^{2}(\sqrt{\alpha}\, x_{i}) \, dx_{i}\Bigg] \nonumber \\
& = \left(\frac{\alpha^{3}}{\pi}\right)^{\frac{1}{2}} \Bigg[\Pi_{i=1}^{D} \frac{1}{2^{n_{i}}n_{i}!}\int_{-\infty}^{\infty} x_{i}^{2}\,e^{-\alpha x_{i}^{2}}H_{n_{i}}^{2}(\sqrt{\alpha}\, x_{i}) \, dx_{i}\Bigg]\nonumber \\
%&\hspace{-2cm} = \underbrace{\left(\frac{\alpha^{3}}{\pi}\right)^{\frac{1}{2}}\frac{1}{\alpha^{\frac{3}{2}}}}_{=\frac{1}{\sqrt{\pi}}} \Bigg[ \frac{1}{2^{n_{1}}n_{1}!}\int_{-\infty}^{\infty} y_{1}^{2}\,e^{-y_{1}^{2}}H_{n_{1}}^{2}(y_{1}) \, dy_{1} + \ldots  +\frac{1}{2^{n_{D}}n_{D}!}  \int_{-\infty}^{\infty} y_{D}^{2}\, e^{-y_{D}^{2}}H_{n_{D}}^{2}(y_{D}) \, dy_{D} \Bigg]\nonumber \\
%&= \frac{1}{\sqrt{\pi}} \Bigg[\frac{1}{2^{n_{1}}n_{1}!}2^{n_{1}}n_{1}!\sqrt{\pi}\left(n_{1}+\frac{1}{2}\right) + \ldots + \frac{1}{2^{n_{D}}n_{D}!}2^{n_{D}}n_{D}!\sqrt{\pi}\left(n_{D}+\frac{1}{2}\right) \Bigg]\nonumber \\
%&= \Bigg[ n_{1} +\frac{1}{2} +  \ldots + n_{D} +\frac{1}{2} \Bigg]\nonumber \\
&=  \Bigg( n_{1} +  \ldots + n_{D} +\frac{D}{2} \Bigg) = \Bigg( N +\frac{D}{2} \Bigg).
\end{align}
Now, let us tackle the third integral $\mathcal{I}_{3}^{(D)}$,
\begin{align*}
\mathcal{I}_{3}^{(D)} &=-\mathcal{N}^{2}\int_{\mathbb{R}^{D}} e^{-\alpha(x_{1}^{2}+x_{2}^{2}+\ldots+x_{D}^{2})}H_{n_{1}}^{2}(\sqrt{\alpha}\, x_{1})\cdots H_{n_{D}}^{2}(\sqrt{\alpha}\, x_{D})\nonumber \\
&\hspace{1.5cm}  \times\log\left[H_{n_{1}}^{2}(\sqrt{\alpha}\, x_{1})\cdots H_{n_{D}}^{2}(\sqrt{\alpha}\, x_{D}) \right]\, dx_{1}\cdots dx_{D} \nonumber \\
%&= -\mathcal{N}^{2}\int_{\mathbb{R}^{D}} e^{-\alpha(x_{1}^{2}+x_{2}^{2}+\ldots+x_{D}^{2})}H_{n_{1}}^{2}(\sqrt{\alpha}\, x_{1})\cdots H_{n_{D}}^{2}(\sqrt{\alpha}\, x_{D})\nonumber \\
%&\hspace{1.5cm}  \times \left[ \log H_{n_{1}}^{2}(\sqrt{\alpha}\, x_{1}) +\ldots + \log H_{n_{D}}^{2}(\sqrt{\alpha}\, x_{D}) \right]\, dx_{1}\cdots dx_{D} \nonumber \\
%&= -\mathcal{N}^{2}\Bigg[ \int_{\mathbb{R}^{D}}e^{-\alpha(x_{1}^{2}+x_{2}^{2}+\ldots+x_{D}^{2})}H_{n_{1}}^{2}(\sqrt{\alpha}\, x_{1})\cdots H_{n_{D}}^{2}(\sqrt{\alpha}\, x_{D}) \log H_{n_{1}}^{2}(\sqrt{\alpha}\, x_{1})\,dx_{1}\cdots dx_{D} + \ldots \nonumber \\
%& + \int_{\mathbb{R}^{D}}e^{-\alpha(x_{1}^{2}+x_{2}^{2}+\ldots+x_{D}^{2})}H_{n_{1}}^{2}(\sqrt{\alpha}\, x_{1})\cdots H_{n_{D}}^{2}(\sqrt{\alpha}\, x_{D}) \log H_{n_{D}}^{2}(\sqrt{\alpha}\, x_{1})\,dx_{1}\cdots dx_{D}\Bigg]\nonumber \\
&= -\mathcal{N}^{2}\Bigg[ \int_{-\infty}^{\infty} e^{-\alpha x_{1}^{2}}H_{n_{1}}^{2}(\sqrt{\alpha}\, x_{1})\log H_{n_{1}}^{2}(\sqrt{\alpha}\, x_{1})\,dx_{1}\cdots \int_{-\infty}^{\infty} e^{-\alpha x_{D}^{2}}H_{n_{D}}^{2}(\sqrt{\alpha}\, x_{D})\,dx_{D}   + \ldots \nonumber \\
& + \int_{-\infty}^{\infty} e^{-\alpha x_{1}^{2}}H_{n_{1}}^{2}(\sqrt{\alpha}\, x_{1})\,dx_{1}\cdots \int_{-\infty}^{\infty} e^{-\alpha x_{D}^{2}}H_{n_{D}}^{2}(\sqrt{\alpha}x_{D})\log H_{n_{D}}^{2}(\sqrt{\alpha}\, x_{D})\, \,dx_{D}  \Bigg]\nonumber \\
%&= -\mathcal{N}^{2}\Bigg[\left(\frac{\pi}{\alpha}\right)^{\frac{D-1}{2}}2^{n_{2}+\ldots + n_{D}}n_{2}!\cdots n_{D}! \int_{-\infty}^{\infty} e^{-\alpha x_{1}^{2}}H_{n_{1}}^{2}(\sqrt{\alpha}\, x_{1})\log H_{n_{1}}^{2}(\sqrt{\alpha}\, x_{1})\,dx_{1}   + \ldots \nonumber \\
%& + \left(\frac{\pi}{\alpha}\right)^{\frac{D-1}{2}}2^{n_{1}+\ldots + n_{D-1}}n_{1}!\cdots n_{D-1}! \int_{-\infty}^{\infty} e^{-\alpha x_{D}^{2}}H_{n_{D}}^{2}(\sqrt{\alpha}x_{D})\log H_{n_{D}}^{2}(\sqrt{\alpha}\, x_{D})\, \,dx_{D}  \Bigg]\nonumber \\
%&= -\mathcal{N}^{2}\left(\frac{\pi}{\alpha}\right)^{\frac{D-1}{2}}2^{n_{1}+\ldots + n_{D}}n_{1}!\cdots n_{D}!\Bigg[\frac{1}{2^{n_{1}}n_{1}!} \int_{-\infty}^{\infty} e^{-\alpha x_{1}^{2}}H_{n_{1}}^{2}(\sqrt{\alpha}\, x_{1})\log H_{n_{1}}^{2}(\sqrt{\alpha}\, x_{1})\,dx_{1}   + \ldots \nonumber \\
%& + \frac{1}{2^{n_{D}}n_{D}!} \int_{-\infty}^{\infty} e^{-\alpha x_{D}^{2}}H_{n_{D}}^{2}(\sqrt{\alpha}x_{D})\log H_{n_{D}}^{2}(\sqrt{\alpha}\, x_{D})\, \,dx_{D}  \Bigg]\nonumber \\
%& = -\left(\frac{\alpha}{\pi}\right)^{\frac{1}{2}}\Bigg[\frac{1}{2^{n_{1}}n_{1}!} \int_{-\infty}^{\infty} e^{-\alpha x_{1}^{2}}H_{n_{1}}^{2}(\sqrt{\alpha}\, x_{1})\log H_{n_{1}}^{2}(\sqrt{\alpha}\, x_{1})\,dx_{1}   + \ldots\nonumber \\
%&  + \frac{1}{2^{n_{D}}n_{D}!} \int_{-\infty}^{\infty} e^{-\alpha x_{D}^{2}}H_{n_{D}}^{2}(\sqrt{\alpha}x_{D})\log H_{n_{D}}^{2}(\sqrt{\alpha}\, x_{D})\, \,dx_{D}  \Bigg]\nonumber \\
& = -\frac{1}{\sqrt{\pi}}\Bigg[\Pi_{i=1}^{D}\frac{1}{2^{n_{i}}n_{i}!} \int_{-\infty}^{\infty} e^{-y_{i}^{2}}H_{n_{i}}^{2}(y_{i})\log H_{n_{i}}^{2}(y_{i})\,dy_{i} \Bigg]\nonumber \\
& = -\frac{1}{\sqrt{\pi}}\Bigg[\Pi_{i=1}^{D}\frac{1}{2^{n_{i}}n_{i}!} E_{n_i}(H)\Bigg]
%&\hspace{-3cm} = \frac{1}{\sqrt{\pi}}\Bigg[\frac{1}{2^{n_{1}}n_{1}!} 2^{n_{1}}n_{1}!\sqrt{\pi}\left(n_{1}\gamma - 2\sum_{i=1}^{n_{1}}x_{n_{1},i}^{2} \, {}_2 F_{2}\left(1,1;\frac{3}{2},2;-x_{n_{1},i}^{2} \right)+\sum_{k=1}^{n_{1}}\binom{n_{1}}{k}\frac{(-1)^{k}2^{k}}{k}\sum_{i=1}^{n_{1}}\, {}_1 F_{1}\left(k;\frac{1}{2};-x_{n_{1},i}^{2} \right) \right) \nonumber \\
%& \hspace{-3cm} + \ldots  + \frac{1}{2^{n_{D}}n_{D}!} 2^{n_{D}}n_{D}!\sqrt{\pi}\left(n_{D}\gamma - 2\sum_{i=1}^{n_{D}}x_{n_{D},i}^{2}\, {}_2 F_{2}\left(1,1;\frac{3}{2},2;-x_{n_{D},i}^{2} \right)+\sum_{k=1}^{n_{D}}\binom{n_{D}}{k}\frac{(-1)^{k}2^{k}}{k}\sum_{i=1}^{n_{D}}\, {}_1 F_{1}\left(k;\frac{1}{2};-x_{n_{D},i}^{2} \right) \right) \Bigg]\nonumber \\
%&\hspace{-3cm} = \Bigg[\left(n_{1}\gamma - 2\sum_{i=1}^{n_{1}}x_{n_{1},i}^{2}\, {}_2 F_{2}\left(1,1;\frac{3}{2},2;-x_{n_{1},i}^{2} \right)+\sum_{k=1}^{n_{1}}\binom{n_{1}}{k}\frac{(-1)^{k}2^{k}}{k}\sum_{i=1}^{n_{1}}\, {}_1 F_{1}\left(k;\frac{1}{2};-x_{n_{1},i}^{2} \right) \right) \nonumber \\
%& \hspace{-3cm} + \ldots  + \left(n_{D}\gamma - 2\sum_{i=1}^{n_{D}}x_{n_{D},i}^{2}\, {}_2 F_{2}\left(1,1;\frac{3}{2},2;-x_{n_{D},i}^{2} \right)+\sum_{k=1}^{n_{D}}\binom{n_{D}}{k}\frac{(-1)^{k}2^{k}}{k}\sum_{i=1}^{n_{D}}\, {}_1 F_{1}\left(k;\frac{1}{2};-x_{n_{D},i}^{2} \right) \right) \Bigg],
\end{align*}
where $E_{n_i}(H)$ is the entropy-like integral functional of the orthogonal Hermite polynomials previously found by various authors \cite{jordi,wolfram} as
\begin{equation*}
E_{n}(H) \equiv \int_{0}^{\infty} [H_{n}(x)]^{2}\log[H_{n}(x)]^{2}e^{-x^{2}}\, dx = 2^{n}n!\sqrt{\pi}\log(2^{2n})-2\sum_{k=1}^{n}V_{n}(x_{n,k})\, .
\end{equation*}
The symbol $\{x_{n,k}, k=1, ...,n\}$ denotes the roots of the Hermite polynomial $H_n(x)$, and $V_{n}(x)$ is known as the logarithmic potential of the Hermite polynomial $H_n(x)$ which is given by
\[
V_{n}(x)=2^{n}n!\sqrt{\pi}\left[\log 2+\frac{\gamma}{2}-x^{2}\, {}_2 F_{2}\left(1,1;\frac{3}{2},2;-x^{2} \right)+\frac{1}{2}\sum_{i=1}^{n}\binom{n}{k}\frac{(-1)^{k}2^{k}}{k}\, {}_1 F_{1}\left(1;\frac{1}{2};-x^{2} \right) \right],
\] 
where $\gamma$ is the Euler constant and $_1F_1$ and $_2F_2$ denote the well known hypergeometric functions \cite{olver}. Then, rewriting $\mathcal{I}_{3}^{(D)} $ more appropriately one has that
\begin{align}
\label{I31}
\mathcal{I}_{3}^{(D)} &= N\gamma  - 2\left( \sum_{i=1}^{n_{1}}x_{n_{1},i}^{2}\, {}_2 F_{2}\left(1,1;\frac{3}{2},2;-x_{n_{1},i}^{2} \right) + \ldots + \sum_{i=1}^{n_{D}}x_{n_{D},i}^{2}\, {}_2 F_{2}\left(1,1;\frac{3}{2},2;-x_{n_{D},i}^{2} \right) \right) \nonumber \\
&+\sum_{k=1}^{n_{1}}\binom{n_{1}}{k}\frac{(-1)^{k}2^{k}}{k}\sum_{i=1}^{n_{1}}\, {}_1 F_{1}\left(k;\frac{1}{2};-x_{n_{1},i}^{2} \right)  + \ldots   +\sum_{k=1}^{n_{D}}\binom{n_{D}}{k}\frac{(-1)^{k}2^{k}}{k}\sum_{i=1}^{n_{D}}\, {}_1 F_{1}\left(k;\frac{1}{2};-x_{n_{D},i}^{2} \right) .
\end{align}
%\end{itemize}
Then, gathering into  (\ref{cchose1}) Eqs. (\ref{I11}), (\ref{I21}) and (\ref{I31}) we have that the exact Shannon entropy of the $D$-dimensional harmonic oscillator in position space is given by
\begin{align}
\label{eseho}
S[\rho_{\{n_{i} \}}] 
%&=  \mathcal{I}_{1}^{(D)} +   \mathcal{I}_{2}^{(D)}  +  \mathcal{I}_{3}^{(D)} \nonumber\\
%&= -N\log 2 -\sum_{i=1}^{D} \log (n_{i}!) - \frac{D}{2} \log \left(\frac{\pi}{\alpha}\right)+ \Bigg( N +\frac{D}{2} \Bigg)  -N\gamma\nonumber\\
&= N\log (2e^{1+\gamma}) +\sum_{i=1}^{D} \log (n_{i}!) + \frac{D}{2} \log \left(\frac{e\,\pi}{\alpha}\right) \nonumber\\
&   - 2\left( \sum_{i=1}^{n_{1}}x_{n_{1},i}^{2}\, {}_2 F_{2}\left(1,1;\frac{3}{2},2;-x_{n_{1},i}^{2} \right) + \ldots + \sum_{i=1}^{n_{D}}x_{n_{D},i}^{2}\, {}_2 F_{2}\left(1,1;\frac{3}{2},2;-x_{n_{D},i}^{2} \right) \right) \nonumber \\
&+\sum_{k=1}^{n_{1}}\binom{n_{1}}{k}\frac{(-1)^{k}2^{k}}{k}\sum_{i=1}^{n_{1}}\, {}_1 F_{1}\left(k;\frac{1}{2};-x_{n_{1},i}^{2} \right)  + \ldots   +\sum_{k=1}^{n_{D}}\binom{n_{D}}{k}\frac{(-1)^{k}2^{k}}{k}\sum_{i=1}^{n_{D}}\, {}_1 F_{1}\left(k;\frac{1}{2};-x_{n_{D},i}^{2} \right)  ,
\end{align}
where $N=n_{1}+\ldots+n_{D}$. We observe that the physical Shannon entropy of the multidimensional harmonic oscillator depends on the position-space dimensionality and the location of the roots of the Hermite polynomials which control the position wavefunctions of the system.
Note that for the ground state ${\{n_i= 0, i = 1,..., D\}}$ one has the value 
$$S[\rho_{\{0,...0\}}] = \frac{D}{2} \log \left(\frac{e\,\pi}{\alpha}\right) $$
(since all the involved sums boil down to zero \cite{adegoke16bis}) and in the particular case $D=1$ one obtains
\begin{align}
S[\rho_{n_1}] &= \log(2^{n_1}n_1!\sqrt{\pi})+n_1+\frac{1}{2}+n_1\gamma-2 \sum_{i=1}^{n_1}x_{n_1,i}^{2}\, {}_2 F_{2}\left(1,1;\frac{3}{2},2;-x_{n_1,i}^{2} \right) \nonumber \\ 
&+ \sum_{k=1}^{n_1}\binom{n_1}{k}\frac{(-1)^{k}2^{k}}{k}\sum_{i=1}^{n_1}\, {}_1 F_{1}\left(k;\frac{1}{2};-x_{n_1,i}^{2} \right),
\end{align}
which is the position Shannon entropy for the one-dimensional harmonic oscillator previously obtained \cite{jordi} as expected. Moreover, in the attached figure we give for illustrative purposes the position Shannon entropy for the states $\{n_i=0, i = 1,\ldots, D\}, \{n_1=1; n_i=0, i = 2,\ldots, D\}, \{n_i=1,i = 1,\ldots, D-1; n_D =0\}$ and $ \{n_i=1, i = 1,\ldots, D\}$ at various dimensionalities. Therein, we practically observe a monotonically increasing entropic behavior when both the dimensionality and the population of the states are increasing.
\begin{figure}[h!]
\includegraphics[scale=1]{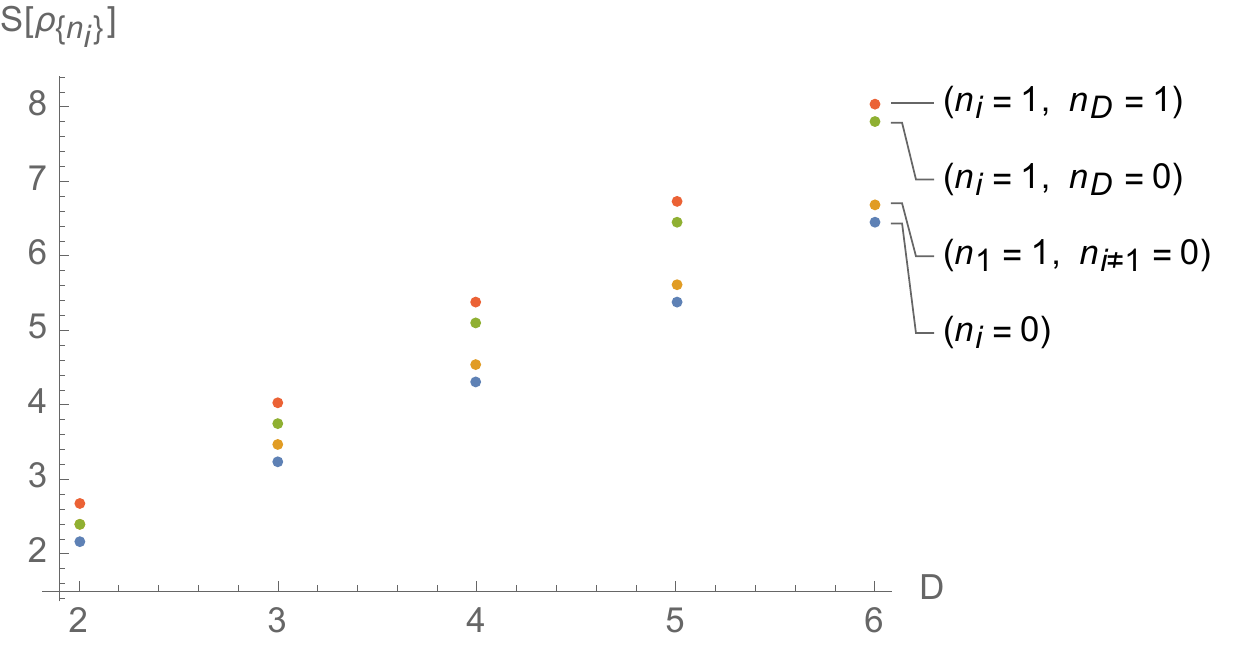}
\centering
\caption{Shannon entropy of the $D$-dimensional harmonic oscillator for four configuration states in terms of the spatial dimension $D$.}
\end{figure}
%As can be observed, this behaviour es monotonically increasing for each state; also the Shannon entropy increases with the principal quantum number, $n$.  
Finally, keeping in mind the trivial relation between the position and momentum probability densities for the harmonic system as explicitly mentioned above, one obtains the following expression for the Shannon entropy in momentum space
\begin{align}
\label{esehom}
S[\gamma_{\{n_{i} \}}]
&= N\log (2e^{1+\gamma}) +\sum_{i=1}^{D} \log (n_{i}!) + \frac{D}{2} \log (e\,\pi\alpha) \nonumber\\
&   - 2\left( \sum_{i=1}^{n_{1}}x_{n_{1},i}^{2}\, {}_2 F_{2}\left(1,1;\frac{3}{2},2;-x_{n_{1},i}^{2} \right) + \ldots + \sum_{i=1}^{n_{D}}x_{n_{D},i}^{2}\, {}_2 F_{2}\left(1,1;\frac{3}{2},2;-x_{n_{D},i}^{2} \right) \right) \nonumber \\
&+\sum_{k=1}^{n_{1}}\binom{n_{1}}{k}\frac{(-1)^{k}2^{k}}{k}\sum_{i=1}^{n_{1}}\, {}_1 F_{1}\left(k;\frac{1}{2};-x_{n_{1},i}^{2} \right)  + \ldots   +\sum_{k=1}^{n_{D}}\binom{n_{D}}{k}\frac{(-1)^{k}2^{k}}{k}\sum_{i=1}^{n_{D}}\, {}_1 F_{1}\left(k;\frac{1}{2};-x_{n_{D},i}^{2} \right)\, .
\end{align}
Thus, the Shannon uncertainty sum gives
\begin{align}
\label{susho}
S[\rho_{\{n_{i} \}}] + S[\gamma_{\{n_{i} \}}] &= N\log \left(4e^{2(1+\gamma)}\right) +\sum_{i=1}^{D} \log (n_{i}!)^{2} + D \log (e\,\pi) \nonumber\\
&   - 4\left( \sum_{i=1}^{n_{1}}x_{n_{1},i}^{2}\, {}_2 F_{2}\left(1,1;\frac{3}{2},2;-x_{n_{1},i}^{2} \right) + \ldots + \sum_{i=1}^{n_{D}}x_{n_{D},i}^{2}\, {}_2 F_{2}\left(1,1;\frac{3}{2},2;-x_{n_{D},i}^{2} \right) \right) \nonumber \\
&\hspace{-2cm} +2\left[\sum_{k=1}^{n_{1}}\binom{n_{1}}{k}\frac{(-1)^{k}2^{k}}{k}\sum_{i=1}^{n_{1}}\, {}_1 F_{1}\left(k;\frac{1}{2};-x_{n_{1},i}^{2} \right)  + \ldots   +\sum_{k=1}^{n_{D}}\binom{n_{D}}{k}\frac{(-1)^{k}2^{k}}{k}\sum_{i=1}^{n_{D}}\, {}_1 F_{1}\left(k;\frac{1}{2};-x_{n_{D},i}^{2} \right)\right]  ,
\end{align}
which does not depend on the oscillator parameter $\alpha$ as expected because the harmonic potential is an homogeneous potential. Moreover, for the ground state ${\{n_i= 0, i = 1,..., D\}}$ one has the value
$S[\rho_{\{0,...0\}}] + S[\gamma_{\{0,...0\}}] = D \log (e\,\pi)$, which saturates the well-known entropic uncertainty relation of Bialynicki-Birula and Mycielski \cite{bbm}.

%\subsection{Hydrogenic systems}

\section{Conclusions}

Space dimensionality is known to have a direct influence on the chemical and physical properties of the quantum systems, basically because their wavefunctions strongly depend on it. Moreover it has been recently shown that space dimensionality is a physico-technological resource in quantum information and computation \cite{krenn,plastino,zeilinger,brunner} and nanotechnology \cite{harrison}. Presently there is an increasing interest on the dimensional dependence of the entropic properties for the stationary states of the multidimensional quantum systems \cite{sen,dehesa_sen12}; it is known that they do not depend on the energy eigenvalues but on the eigenfunctions. Up until now the R\'enyi, Shannon and Tsallis entropies have been computed for the stationary hydrogenic and harmonic states only at the two extreme situations, the high-energy (i.e., Rydberg) \cite{tor2016b,aptekarev2016,dehesa2017} and high-dimensional (i.e., pseudoclassical) \cite{puertas2017,tor2017b} states,  by means of modern asymptotical techniques of Laguerre and Gegenbauer polynomials. 
In this work we have analytically determined the uncertainty measure of Shannon type for the $D$-dimensional harmonic systems from first principles; that is, in terms of the spatial dimension $D$, the oscillator strength $\alpha$, as well as the hyperquantum numbers, $\{n_{i}\}_{i=1}^{D}$, which characterize the corresponding state's wavefunction. This has been possible because the harmonic eigenfunctions in Cartesian coordinates can be expressed in terms of a product of Hermite polynomials and exponentials.
It remains as an open problem the exact calculation of the Shannon entropy for all the discrete stationary states of the main prototype of the multidimnsional Coulomb systems, $D$-dimensional hydrogenic systems .\\

\section*{Acknowledgments}
This work has been partially supported by the Projects FQM-207 and FQM-7276 of the Junta de Andaluc\'ia and the MINECO-FEDER grants FIS2017-89349P and FIS2014-59311-P.

\end{document}